\documentclass[aps,pre,twocolumn,groupaddress]{revtex4}
\usepackage{amsmath}
\usepackage{amsfonts}
\usepackage{amssymb}
\usepackage{epsfig}
\usepackage{graphicx}

\begin{document}

\title{Probing a liquid to glass transition in equilibrium}

\author{Walter Kob$^*$}
\affiliation{Laboratoire Charles Coulomb, UMR 5221, CNRS and Universit\'e
Montpellier 2, Montpellier, France}

\author{Ludovic Berthier}
\affiliation{Laboratoire Charles Coulomb, UMR 5221, CNRS and Universit\'e
Montpellier 2, Montpellier, France}

\begin{abstract}
We use computer simulations to investigate the static properties of a
simple glass-forming fluid in which the positions of 
a finite fraction of the particles has been
frozen in. By probing the equilibrium
distribution of the overlap between independent
configurations of the liquid, we find strong evidence
that this random pinning induces a glass transition. At low temperatures, 
our numerical findings are consistent with the existence of
a random first order phase transition rounded by finite size effects.

\end{abstract}


\maketitle

Experiments allow to measure the increase of the viscosity of
supercooled liquids approaching the glass transition over more than 15
orders of magnitude. Despite this large range, there is at present
no consensus whether this phenomenon is controlled by an underlying
phase transition, or whether relaxation times progressively increase
down to zero temperature~\cite{book_binder_kob}. There is not even
agreement on the microscopic mechanisms at work, despite the fact that
glassy dynamics is commonly observed in a large variety of materials,
from simple liquids, network forming liquids, to soft and biological
matter~\cite{rmp_berthier_biroli_11}.  Because supercooled liquids
inevitably fall out of equilibrium at the experimental glass temperature
$T_g$, amorphous solids form without encountering any
singularity.

The existence of a thermodynamic transition at a finite temperature
$T_K<T_g$ is highly debated~\cite{rmp_berthier_biroli_11}.
Such a singularity was discussed long ago by Kauzmann~\cite{kauzmann_48}
who pointed out that experimental data for the configurational entropy
extrapolate to zero at $T_K$, the ``Kauzmann temperature''.  Physically,
this would imply that for $T > T_K$ there exist exponentially many (in
the number of particles) equilibrium states (neglecting vibrations),
whereas below $T_K$ this number becomes sub-exponential and the
system forms an ``ideal'' glass.  This view is supported by the
analysis of the temperature dependence of the relaxation dynamics,
which seems to be consistent with the existence of a temperature
$T_0>0$ at which the relaxation time diverges~\cite{richert_98}. This
makes sense if thermodynamic and dynamical singularities occur at the
same temperature, $T_0 \approx T_K$, but so far $T_0$ and $T_K$ have
been determined only by empirical fitting formula and uncontrolled
extrapolations of the behavior of macroscopic observables. These
procedures evidently leave much room for debates on the existence
of such a transition~\cite{tanaka_03,hecksher_08}.

The existence of a true phase transition is appealing from a
theoretical perspective, since a number of analytical results obtained
within various approximate schemes or idealized limits suggest the
existence of a ``random first order transition'' (RFOT) occurring
in glass-forming materials~\cite{kirkpatrick_89,mezard_99}. This
transition results from the temperature evolution of a complex free
energy landscape characterized by a large number of metastable states.
It provides an exact realization of Kauzmann's entropy crisis.  However,
the mean-field character of the RFOT, the technical difficulties
posed by finite-dimensional fluctuations~\cite{biroli_12}, and the
indirect connections between theory and experiments explain why
RFOT is only one among several theoretical scenarios for the glass
transition~\cite{rmp_berthier_biroli_11}.

In the following we demonstrate that it is indeed possible to perform
a detailed {\it equilibrium} study of a liquid-glass transition whose
nature is equivalent to the one possibly occurring at $T_K$ in bulk
glass-formers~\cite{cammarota_12a,berthier_12}. This allows us to probe
numerically, for the first time, the nature of the phase transition and
the microscopic properties of the glass phase at thermal equilibrium. In
particular, we show that glass formation can be located without using
uncontrolled extrapolations, and that equilibrium glasses can be produced.

We consider a 50:50 binary mixture of harmonic
spheres~\cite{berthier_09,kob_12} of diameter ratio 1.4 at constant
density $\rho=0.675$, which we study using molecular dynamics (see
SM for methodological details). This quasi-hard sphere system has an
onset temperature around $T_{\rm on} \approx 10$, and a mode-coupling
temperature $T_{\rm mct} \approx 5.2$~\cite{kob_12}.  All numbers are
expressed in appropriate reduced units (see SM). To efficiently sample
the equilibrium thermodynamic properties of the system, we use replica
exchange molecular dynamics~\cite{lyubartsev_92}.  We carefully ensure
thermalization and equilibrium sampling by requiring that all particles
move several times across the entire simulation box for each state point
and that all replicas properly explore the configuration space.

Our central idea is to induce a glass transition 
by increasing the strength of a random pinning field in a
thermalized dense liquid~\cite{cammarota_12a,berthier_12}.
We first equilibrate
the liquid at a given temperature $T$, before freezing permanently
the position of a certain set of particles. Among various other
pinning geometries~\cite{berthier_12,scheidler_04,biroli_08}, here we
randomly select a finite concentration of particles, $c$.  Working at
constant total number density $\rho$, we then study the equilibrium
properties of the model in the $(c,T)$ phase diagram. For each
state point, we average the results over independent realizations
of the random pinning field~\cite{scheidler_04}.  Intuitively,
pinned particles constrain the available phase space of the
remaining fluid particles, thus impacting their static and dynamic
properties. Indeed, we find that the dynamics slows down dramatically
with $c$ at constant $T$, see Fig.~SM1, in agreement with earlier
studies~\cite{berthier_12,kim_03,krakoviack_11,karmakar_12a}. 

Freezing particles at the positions they occupy at equilibrium
is a key ingredient to this study, since this does not perturb
the interaction part of the Hamiltonian, but only constrains the
available phase space.  Indeed, any ensemble averaged correlation
function measured in the pinned liquid takes the same value as
in the bulk system at $c=0$~\cite{scheidler_04,krakoviack_10}.
Theoretical studies have recently revealed two additional
consequences~\cite{cammarota_12a,cammarota_12b}. First, increasing
$c$ affects the configurational space in the same way as decreasing
temperature does for the bulk liquid~\cite{cammarota_12a}.  In particular,
the RFOT occurring at $T_K^0$ for $c=0$ becomes a transition line
$T_K(c) \geq T_K^0$ with equivalent properties. Second, the pinning
field only induces a glass transition if temperature is low enough
so that metastable states and glassy dynamics are well-developed
at $c=0$.  The glass transition line $T_K(c)$ ends at a second
order critical point~\cite{cammarota_12a}, and no transition exists
for temperatures above this point. Note that other types of pinning
fields~\cite{franz_99,karmakar_12b,hedges_09} produce qualitatively
different results, because they perturb the Hamiltonian and possibly
affect the nature of the glass transition~\cite{cammarota_12b}.

\begin{figure}
\psfig{file=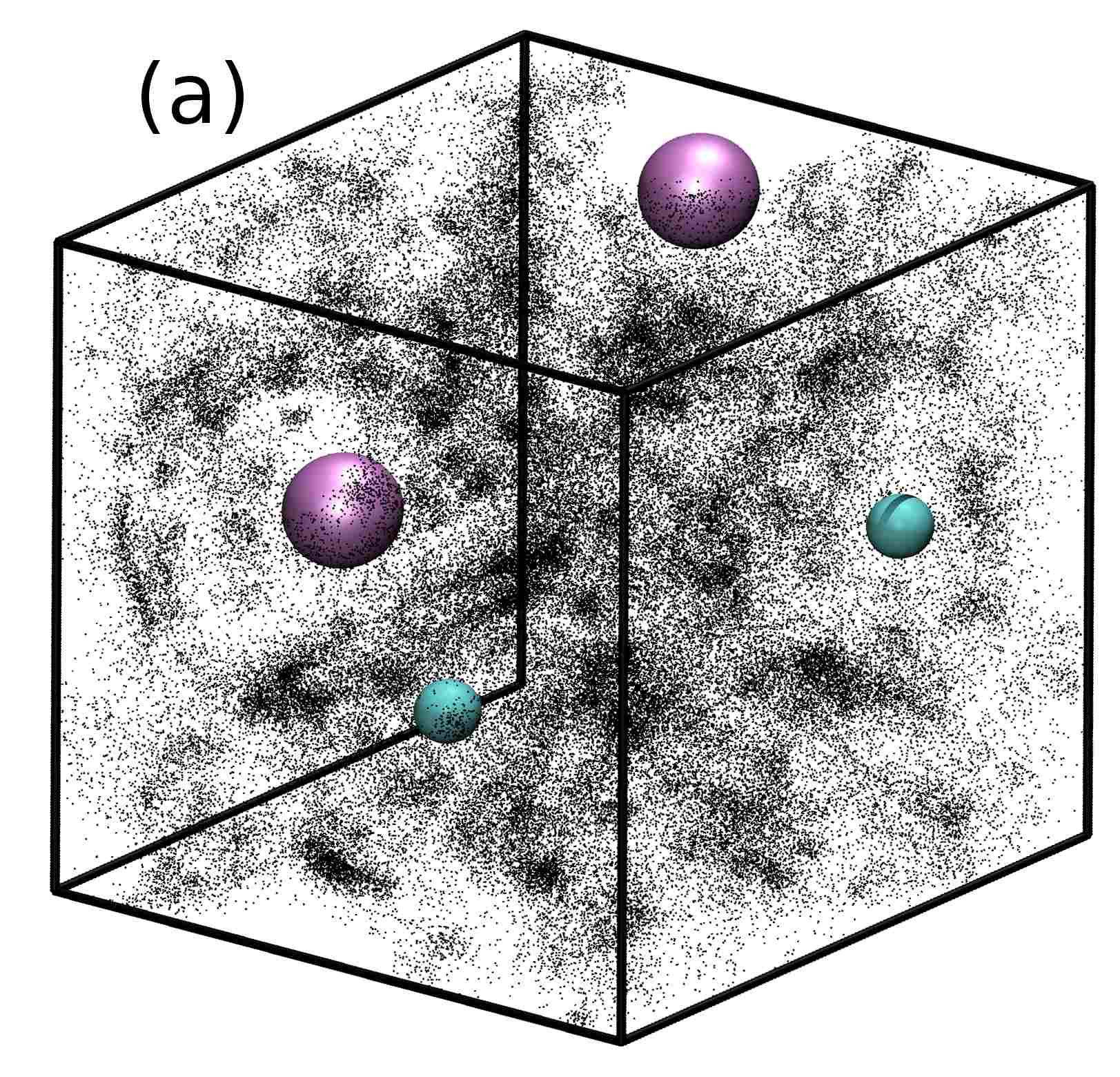,width=4.2cm,clip}
\psfig{file=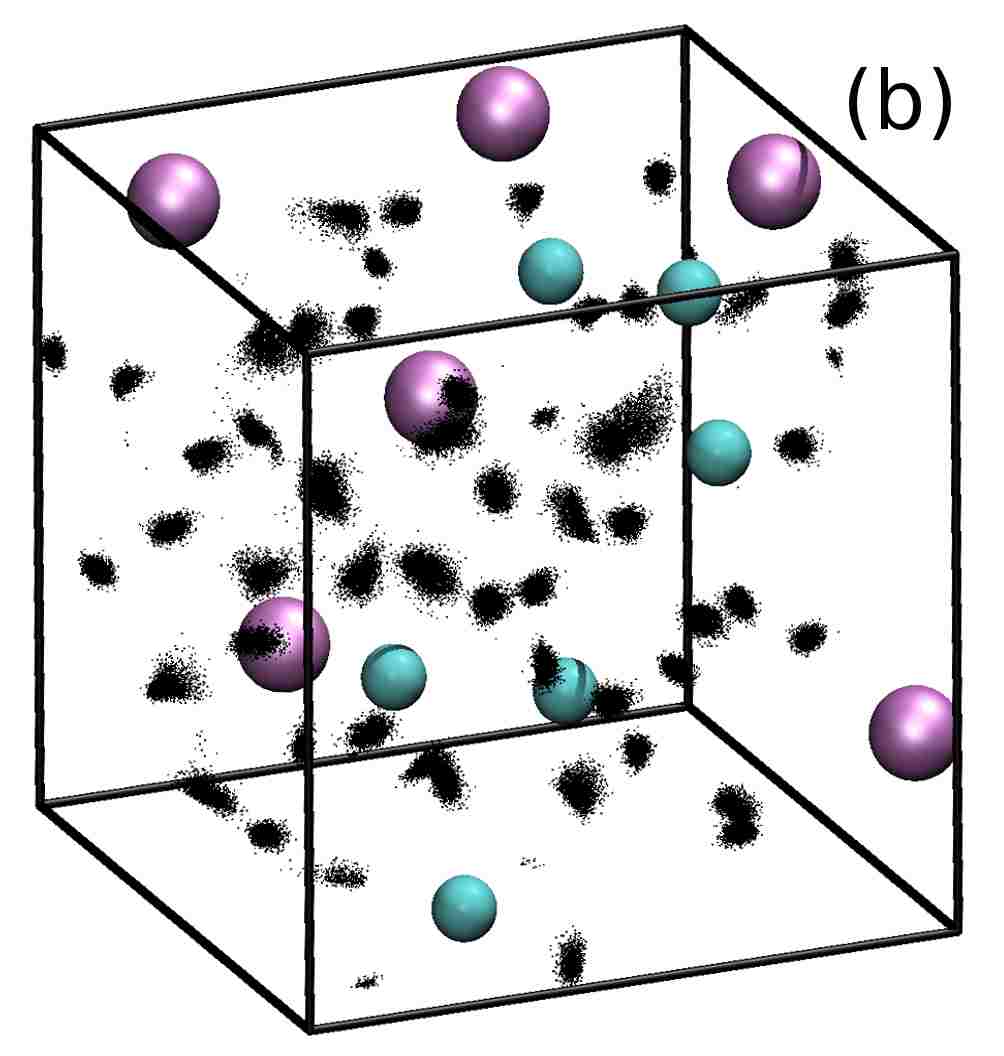,width=3.7cm,clip}
\psfig{file=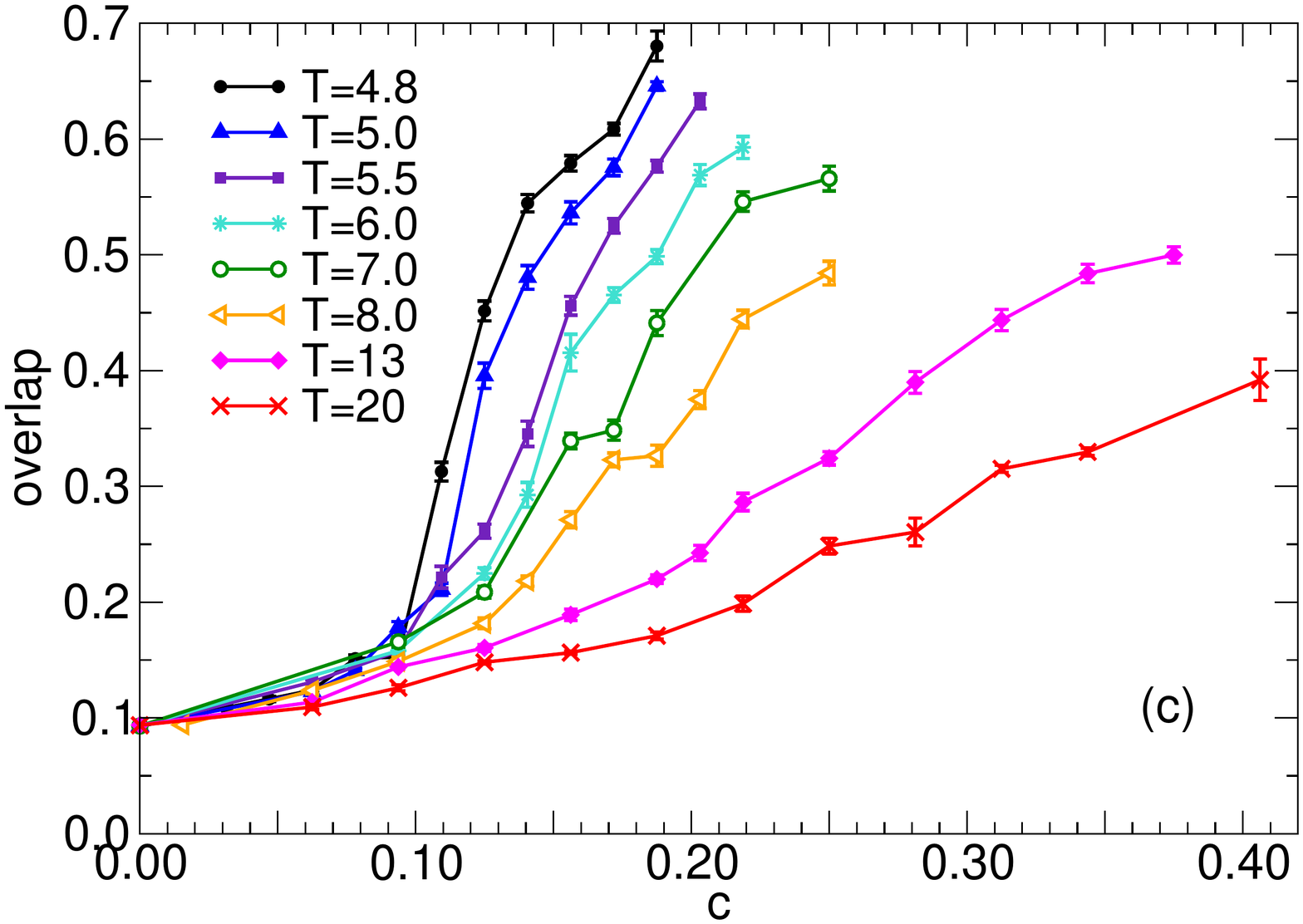,width=8.5cm,clip}
\caption{(a) and (b): Large
spheres represent pinned particles (rescaled in size by a factor 0.5),
small dots are the superposition of the position of fluid particles
obtained from a large number of independent equilibrium configurations
at $T=4.8$ for $c=0.0625$ (a) and $c=0.1875$ (b). The total number of
particles is $N=64$.  Panel (c): Average overlap $\langle q \rangle$
as a function of concentration $c$ of pinned particles 
for $N=64$ and different temperatures.} 
\label{fig1}
\end{figure}

In Fig.~\ref{fig1} we provide a qualitative illustration of our strategy.
We show how an equilibrium glass state is obtained at constant
temperature, $T=4.8$, if the concentration $c$ of pinned particles is
increased. In these snapshots the large spheres are the pinned particles
in a typical realization of the disorder whereas the small black dots
stem from the superposition of a large number of independent, equilibrium
configurations visited by the fluid particles. For $c=0$, the dots look
like mist homogeneously filling the simulation box. For $c=0.0625$,
panel (a), the fluid particles are not much constrained and still have
access to a large number of distinct configurations.  Thus, the dots
form fuzzy clouds. An increase to $c=0.1875$, panel (b), condenses the
dots into well-defined patches, which represent the highly constrained
positions occupied by the free particles.  This qualitative observation
illustrates that increasing $c$ leads to a {\it collective localization}
of the fluid particles. We emphasize that during the simulations, all
fluid particles diffuse and explore the entire simulation box. However,
particles move in such a way that collective density fluctuations
are frozen.  Thus, the system is in a glass state characterized by a
frozen amorphous density profile~\cite{rmp_berthier_biroli_11}, but
which can nevertheless be studied in equilibrium conditions because our
replica exchange simulation algorithm permits single particle diffusion.

We now study in more detail how the system makes the transition between
the fluid and glass states shown in Fig.~\ref{fig1}. The images suggest
that the number of available states, and thus the configurational
entropy, considerably decreases with $c$. A quantitative determination
of the configurational entropy is however difficult and has several
shortcomings~\cite{rmp_berthier_biroli_11}. Therefore we use a microscopic
order parameter to characterize the transition from fuzzy clouds to
small patches seen in Fig.~\ref{fig1}. An appropriate quantity is
the overlap $q_{\alpha \beta}$ measuring the degree of similarity between
two arbitrary configurations $\alpha$ and $\beta$, which has been used in
spin glass models displaying an RFOT~\cite{kirkpatrick_89}. In practice,
we discretize space into small cubic boxes of linear size 0.55,
and define $n_i^{(\alpha)}=1$ if box $i$ in configuration $\alpha$
is occupied by a particle, and $n_i^{(\alpha)}=0$ if not. Then,

\vspace*{-7mm}

\begin{equation}
q_{\alpha \beta} = \frac{1}{N_b} \sum_{i=1}^{N_b} n_i^{(\alpha)} n_i^{(\beta)},
\end{equation}

\noindent
where the sum runs over the $N_b$ boxes which do not contain pinned
particles. By definition $q_{\alpha \alpha}=1$, while a small overlap
is obtained for independent configurations ($q_{\rm rand} \simeq 0.11$
on average for an ensemble of independent configurations).  In the
remainder of the paper, we characterize the transition through a detailed
analysis of the statistical properties of the overlap for a broad range
of control parameters.

In Fig.~\ref{fig1}c, we show the $c$-dependence of the average overlap,
$\langle q \rangle = \langle q_{\alpha \beta} \rangle$, where the brackets
stand for thermal and disorder averages. Above the onset temperature,
$\langle q \rangle$ increases gradually with $c$. For $T \lesssim 8.0$,
the growth remains modest at low $c$, but this initial regime is followed
by a rapid increase in the range $\langle q \rangle \approx 0.25 -
0.4$. Finally, if $T$ is decreased even further this rapid growth occurs
for lower values of $c$ and becomes sharper. At $T=4.8$ it is sufficient
to reach $c \approx 0.11$ to abruptly localize the fluid particles,
$\langle q \rangle > 0.5$, while such a large overlap requires three times
as many pinned particles, $c \approx 0.35$, near the onset temperature.

\begin{figure} 
\psfig{file=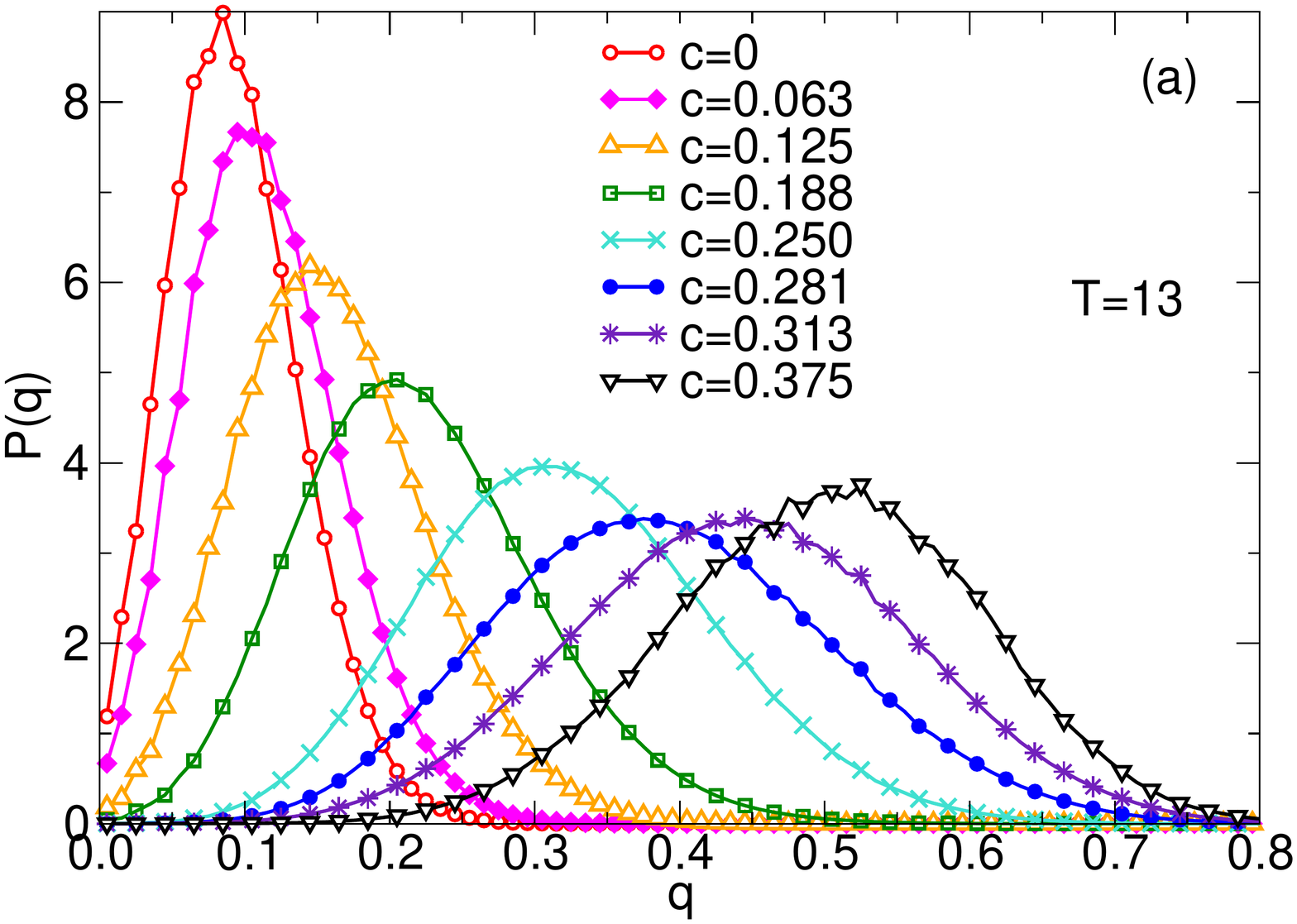,width=8.5cm}
\psfig{file=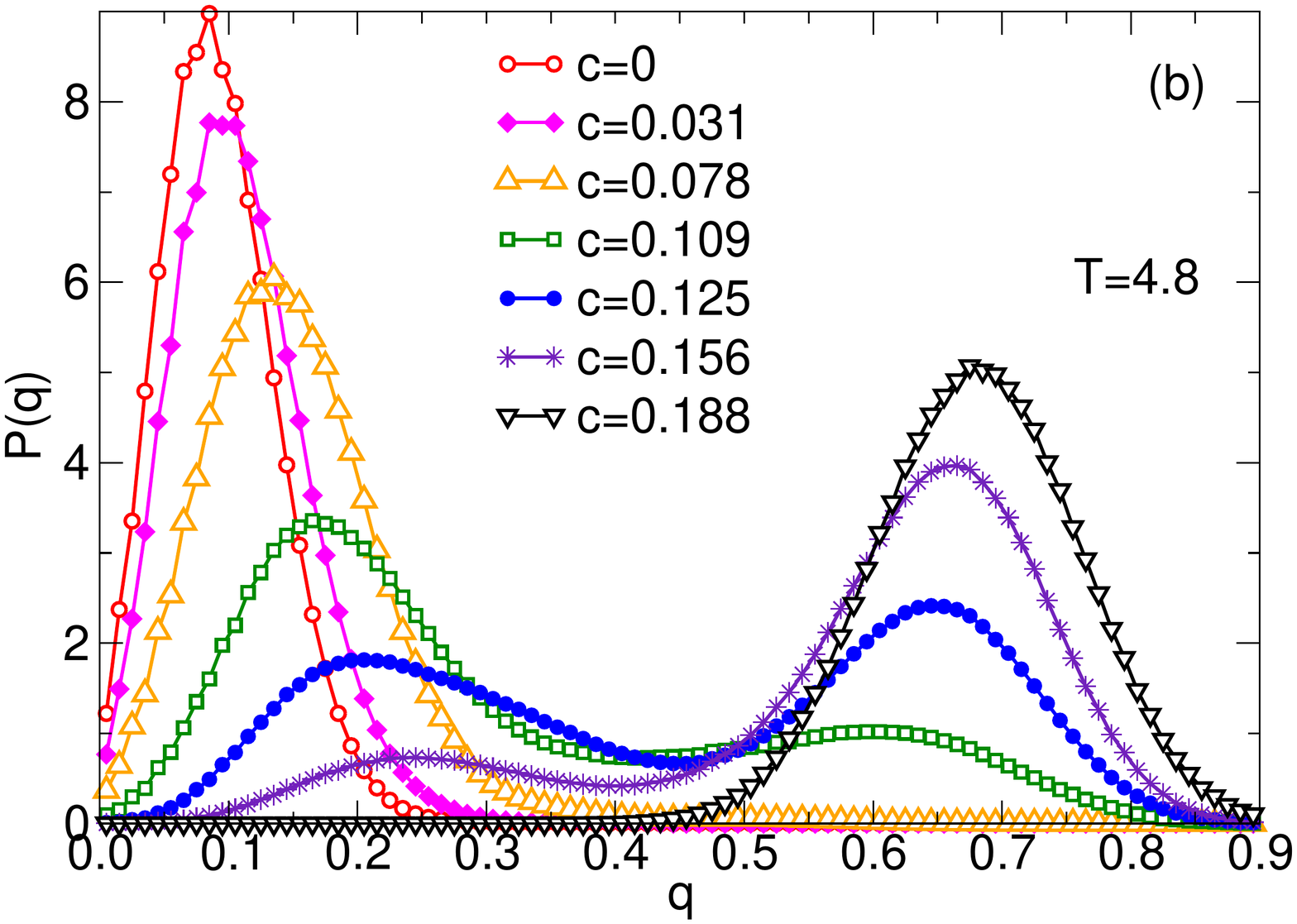,width=8.5cm} 
\caption{Probability distribution of the overlap, Eq.~(\ref{pq}),
for different values of the concentration of pinned particles at high
and low $T$ (panels (a) and (b) respectively). Note the presence of a
double peak structure at low temperatures.}
\label{fig3} 
\end{figure}

These observations show that at low temperatures the glass state is
reached at a sharply defined $c$-value, whereas pinning acts quite
smoothly at high $T$. The $T$-dependence of $\langle q \rangle$
is consistent with the emergence of a discontinuous jump at low
temperatures, $T \lesssim 8.0$, but the relatively small system size
shown in Fig.~\ref{fig1}c, $N=64$, will obviously smear out such
a discontinuity. To overcome this difficulty we have determined the
fluctuations of the overlap and in Fig.~\ref{fig3} we show the overlap
distribution function,

\vspace*{-5mm}

\begin{equation}
P(q) = \langle  \delta(q-q_{\alpha \beta}) \rangle .  
\label{pq}
\end{equation}

At high temperatures, $T=13$ in Fig.~\ref{fig3}a, $P(q)$ evolves smoothly
from distributions peaked at small $q$ for small $c$ to distributions
peaked at large $q$ values at large $c$. This is consistent with $P(q)$
becoming a delta-function for $N \to \infty$ for all $c$, and with the
smooth increase of $\langle q \rangle = \int_0^1 P(q) q dq$  at high
$T$ shown in Fig.~\ref{fig1}c. A qualitatively different behavior is
observed at low $T$, see Fig.~\ref{fig3}b. A narrow peak is still present
for small and large $c$, but $P(q)$ is {\it bimodal} for intermediate
$c$. The presence of two peaks implies that it is equally probable that
two independent thermalized configurations are either very similar or
very different. The former situation is favored at larger $c$ because too
few distinct configurations exist, while the latter holds at small $c$
when the pinning field is not strong enough to prevent the system from
exploring a large configuration space.  Bimodal distributions of the order
parameter can be interpreted as the phase coexistence between the low-$q$
liquid and the high-$q$ glass phases, and the two-peak structure of $P(q)$
is suggestive of a first-order transition for the order parameter $\langle
q \rangle$, rounded by finite size effects.

\begin{figure}
\psfig{file=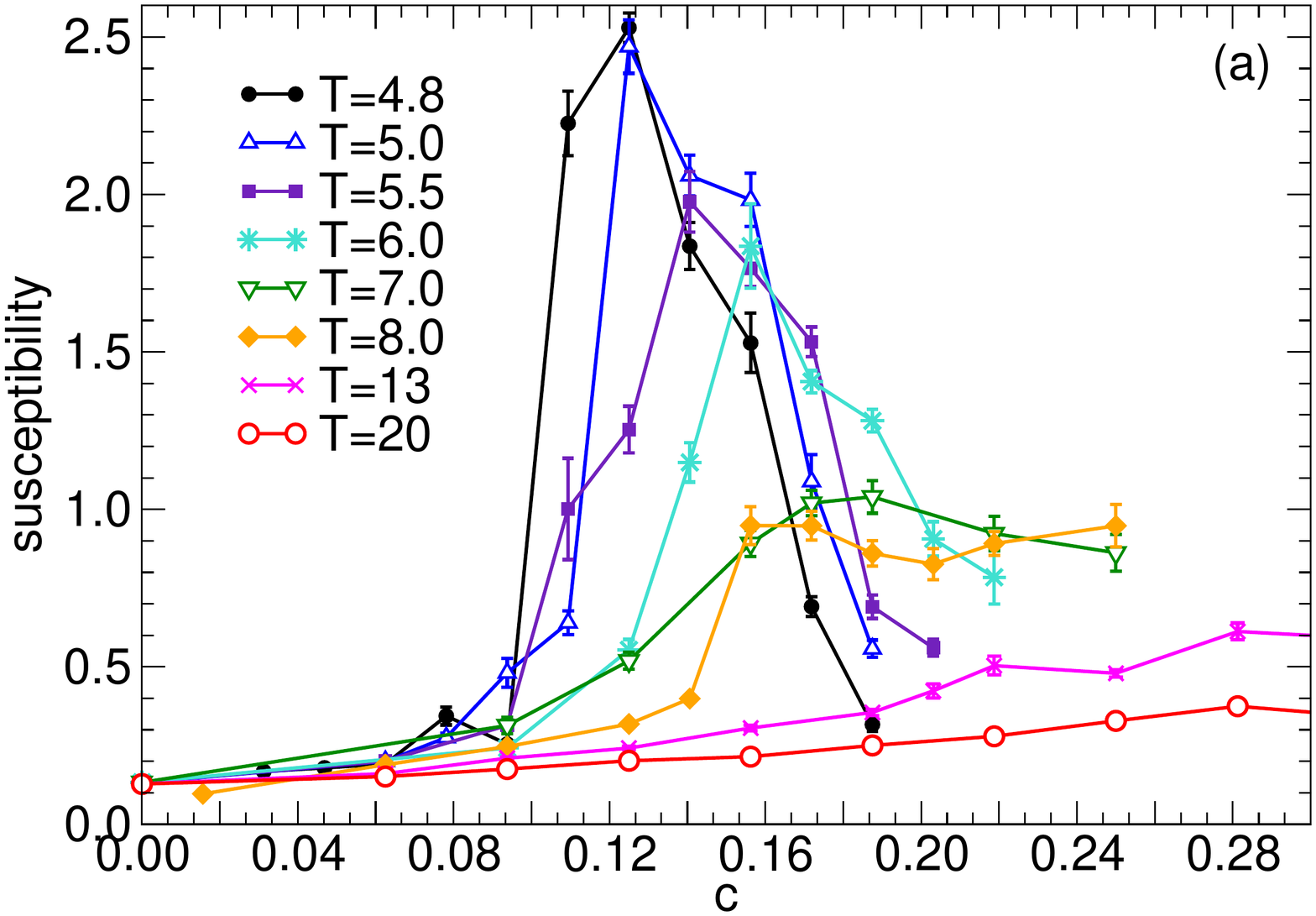,width=8.5cm}
\psfig{file=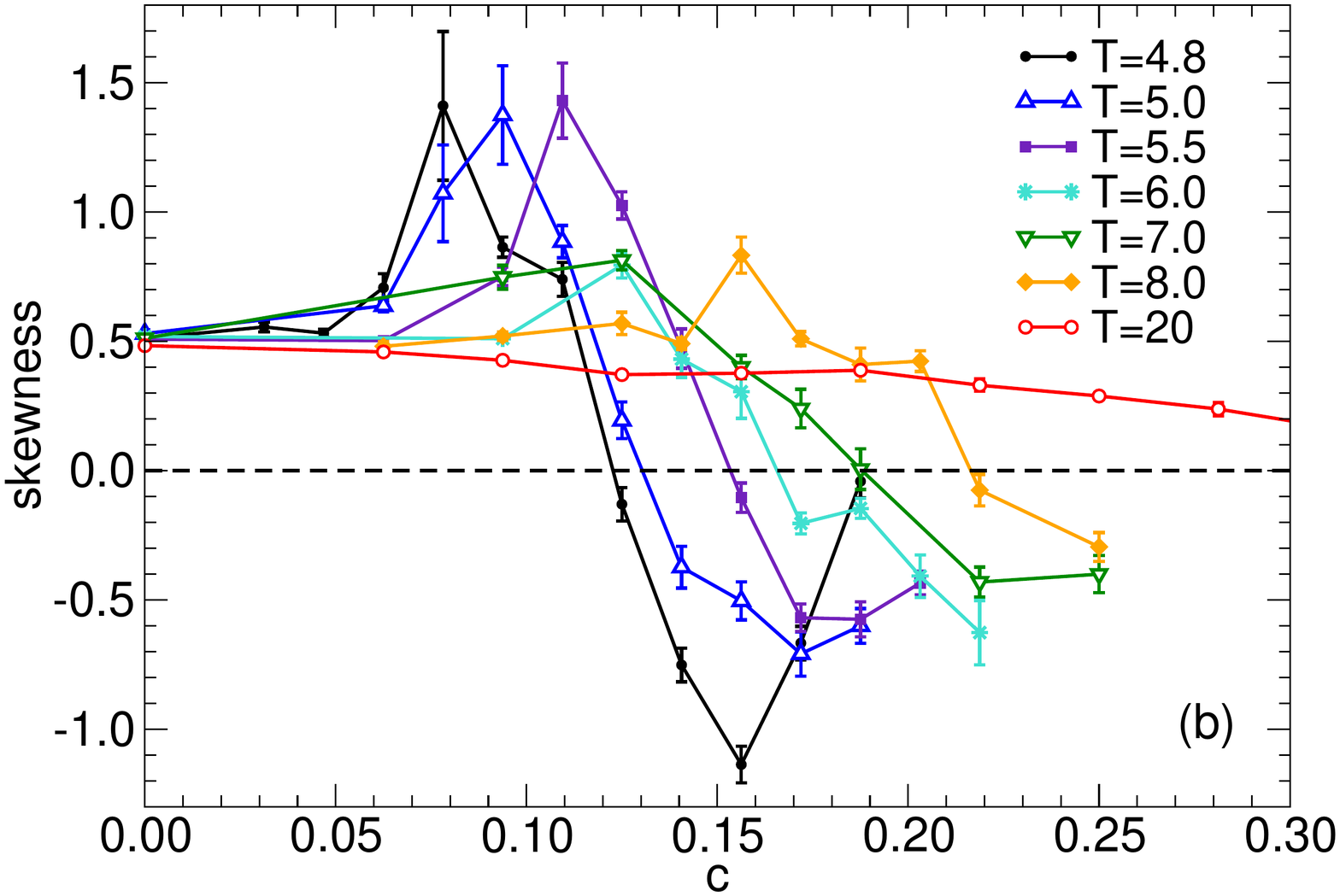,width=8.5cm,clip}
\psfig{file=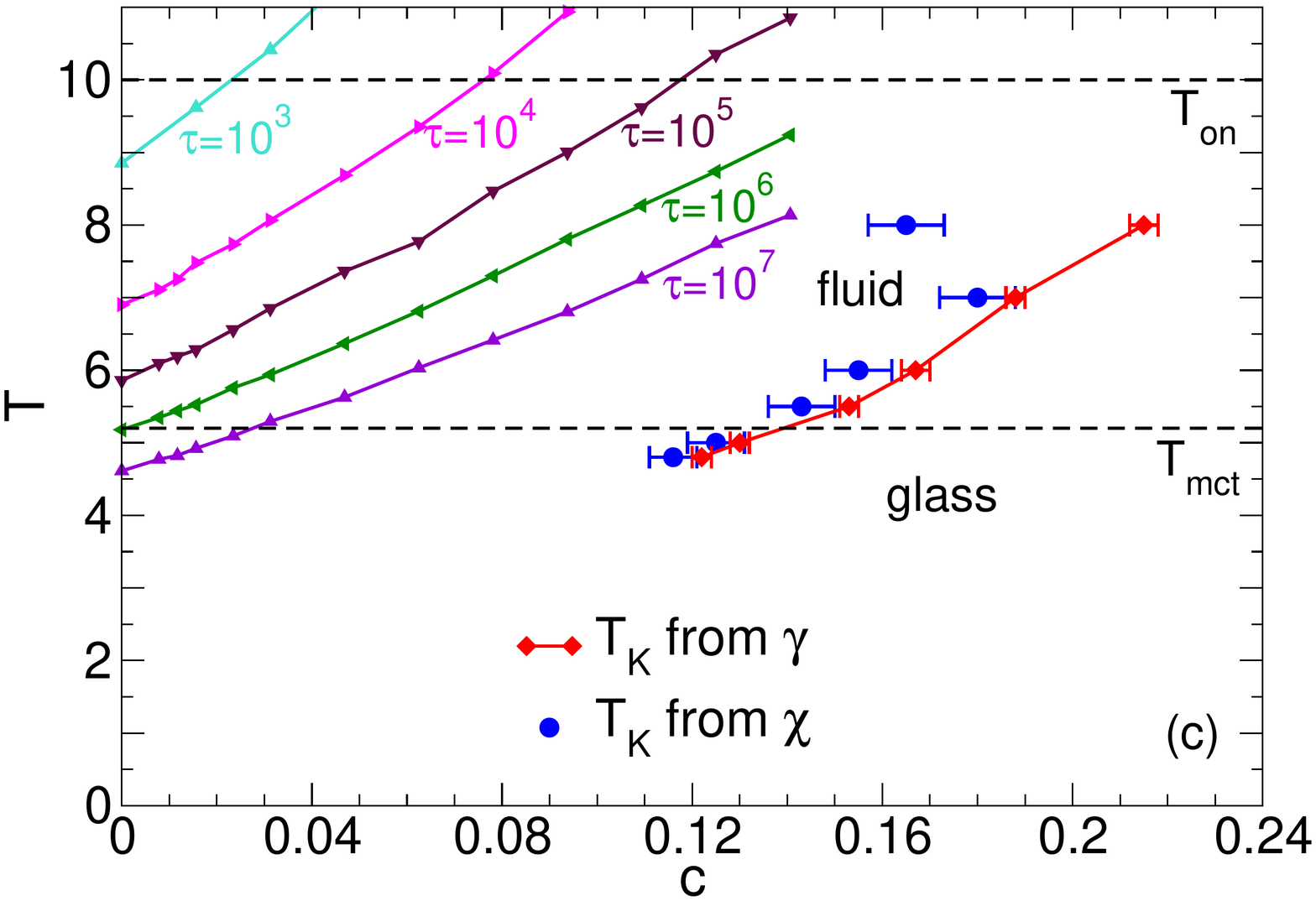,width=8.5cm,clip}
\caption{(a) The static susceptibility $\chi$, Eq.~(\ref{chi}),
as a function of concentration $c$ for different temperatures.  (b)
$c$-dependence of the skewness $\gamma$ for different $T$.  (c) The
equilibrium phase diagram determined from the position of the peak in the
susceptibility $\chi$ and from the root of the skewness $\gamma$ showing
the location of the fluid and glass phases.  Also included are curves
of constant relaxation time as determined from the self-intermediate
scattering function.}
\label{fig4}
\end{figure}

We emphasize that the above conclusions do not result from any kind
of extrapolation, but follow directly from equilibrium measurements of
a microscopic order parameter on {\it both} sides of the transition. To
estimate more precisely the location of this putative transition, 
we have measured
various moments of the distribution $P(q)$. From the second moment
$\langle q^2 \rangle = \int_0^1 P(q) q^2 dq$, we define the static
susceptibility,

\vspace*{-5mm}

\begin{equation}
\chi(c,T) = N (1-c) \left[ \langle q^2 \rangle - \langle q \rangle^2 \right].
\label{chi}
\end{equation}
The curves presented in Fig.~\ref{fig4}a show that at high temperatures
$\chi$ has a mild $c$-dependence, whereas it develops a well-defined
peak for $T \leq 7.0$ whose height and location respectively increases
and shifts to smaller $c$ if $T$ decreases.  The observed growth of the
peak $\chi(c,T)$ at low $T$ is a direct evidence of increasing static
correlations in the bulk system~\cite{berthier_12,jack_12}. The location
of this peak allows to estimate the value of the critical concentration at
the given temperature.  As demonstrated in Fig.~\ref{fig4}b, an even more
accurate location is obtained by considering the skewness $\gamma(c,T)$
of the distribution $P(q)$,

\vspace*{-5mm}

\begin{equation}
\gamma = \frac{ \left\langle \left( q - \langle q \right\rangle \right)^3 
\rangle}{ \left\langle ( q -\langle q \rangle )^2 
\right\rangle^{3/2} },
\end{equation}

\noindent
since $\gamma$ crosses zero when the distribution is symmetric, i.e.
precisely at coexistence. From the location of the peak in $\chi(c,T)$ and
the zero-crossing point of $\gamma(c,T)$, we can locate the fluid-glass
coexistence line, as shown in Fig.~\ref{fig4}c. These two quantitative
estimates become reliable if $T \lesssim 8.0$, i.e. when bimodal
distributions are present, and we see that they agree with each other
within error bars (estimated using the jack-knife method).  In the context
of Fig.~\ref{fig4}c, the existence of a Kauzmann temperature at $c=0$
would rely on extrapolating our finite $c$ transition points (determined
without extrapolation) to the limit $c \to 0$. 

\begin{figure}
\psfig{file=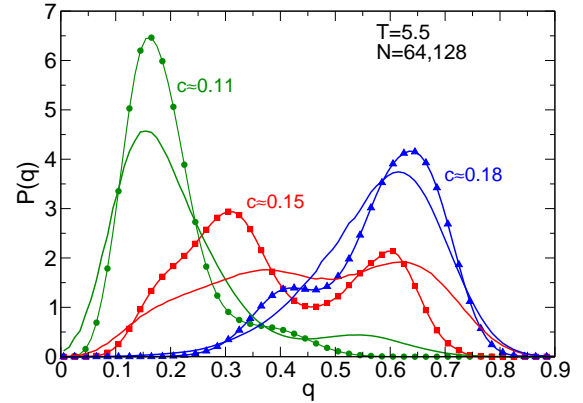,width=8.5cm}
\caption{System size dependence of the overlap distribution function
for $T=5.5$ and three values of $c$ below (green), near (red), and above
(blue) the transition for $N=64$ (lines) and $N=128$ (lines with symbols).
The peak susceptibility increases from $\chi \approx 1.8$ for $N=64$
to $\chi \approx 2.7$ for $N=128$ (data not shown).}
\label{fig5}
\end{figure}

Altogether, our results show that a glass transition is induced 
by random pinning, with nontrivial thermodynamic signatures
occurring at a sharp value of the random pinning field 
that are consistent with the existence of an equilibrium random
first-order transition~\cite{cammarota_12a}. 
However, since our results
have been obtained for a rather small system size, $N=64$, they do not
establish its existence in the thermodynamic limit.  Unfortunately, the
parallel tempering algorithm we use becomes less efficient when $N$ is
increased. While we cannot access very low temperatures for larger $N$,
we can, however, investigate the $N$-dependence in a limited range. Our
results for larger systems confirm that no sharp transition exists
for $T \ge 8.0$, while we obtained clear signs of enhanced bimodality
and static susceptibility when $T < 6.0$, as shown in Fig.~\ref{fig5}
for $T=5.5$. The small dip in $P(q)$ for $N=64$ and intermediate $c$
values becomes more pronounced for $N=128$, while the peaks at low and
large overlaps become sharper. Such system-size dependence is again 
typical of
first-order transitions. A more extensive finite-size scaling analysis
would be useful, but it is at present beyond the computational means.

In contrast to conventional studies of the glass transition, 
our approach allows us to study both liquid and glass phases at
thermal equilibrium, it does not require uncontrolled extrapolations, and
we characterize the glass formation by means of a microscopic order parameter.
Our work thus demonstrates the feasibility of systematic equilibrium
studies of the nature of the liquid-to-glass transition. Our approach
is very general and can easily be applied to any type of glass-former.
Also, by using optical tweezers it will be possible to use this method to
study colloidal glasses and thus to produce experimentally what is often
considered as ``impossible-to-reach ideal glass states''~\cite{chang_08}. 
Therefore our
work opens the door for a new generation of direct, systematic studies of
the nature of the glass state and for the production of novel amorphous
materials.

\acknowledgements 
We thank G. Biroli, 
C. Cammarota, and D. Coslovich for discussions.
This work was partly realized with the support of HPC@LR Center of
Competence in High Performance Computing of Languedoc-Roussillon
(France).  The research leading to these results has received funding
from the European Research Council under the European Union's Seventh
Framework Programme (FP7/2007-2013) / ERC Grant agreement No 306845.
W. Kob acknowledges support from the Institut Universitaire de France.

\newpage

\begin{center}
{\bf Supplementary Material}
\end{center}

\section{Model and details of the simulations}

The model we have been using is a 50:50 binary mixture of $N$ harmonic
spheres~\cite{berthier_09_S}. All particles have the same mass $m$
and particles $i$ and $j$ separated by a distance $r_{ij}$ have an
interaction given by

\begin{equation}
V(r_{ij}) = \frac{\epsilon}{2} (1-(r_{ij}/\sigma_{ij})^2) \quad .
\label{eq1}
\end{equation}

\noindent
Here $\sigma_{11}=1.0$, $\sigma_{12}=1.2$, and $\sigma_{22}=1.4$. We will
use $\sigma_{11}$ as the unit of distance, $\sqrt{m\sigma_{11}/\epsilon}$
as the unit of time and $10^{-4}\epsilon$ as unit of energy, setting the
Boltzmann constant $k_B=1.0$. The density we have been using is 0.6749715.

Before we discuss the details of the procedure how we have pinned
the particles, we recall that pinning particles in a liquid
configuration that has been equilibrated at a temperature $T_{\rm
pin}$ does not perturb the equilibrium properties of the remaining
liquid at the temperature $T_{\rm liq}=T_{\rm pin}$ {\it if one
averages over a sufficiently large number of realisations of pinned
configurations}~\cite{scheidler_04_S,krakoviack_10_S,krakoviack_11_S}. Therefore
it is not necessary to re-equilibrate the pinned system after
the pinning. If instead $T_{\rm pin}$ is different from the
simulation temperature $T_{\rm liq}$ of the liquid the resulting
Kauzmann line $T_K(c,T_{\rm liq})$ will depend on $T_{\rm pin}$ as
well~\cite{cammarota_12b_S}.

To prepare the system at different concentrations $c$ of pinned particles
we have proceeded as follows. First we have used a standard molecular
dynamics simulations to equilibrate a liquid (in the bulk) at the target
temperature $T_{\rm liq}$. For this we have integrated the equations of
motion using the velocity form of the Verlet algorithm with a time step
of 0.02.  Let us call one of the so obtained bulk configurations $C_{\rm
bulk}$.  Subsequently we have to determine which ones of the particles
have to be pinned. Although chosing these pinned particles randomly
from the $C_{\rm bulk}$ configuration is certainly a possibility, such a
choice can lead to systems in which many pinned particles are clustered
or that have relatively large regions in which there are no pinned
particles. These fluctuations make that the transition at the Kauzmann
point becomes smeared out at intermediate temperatures and hence makes
it difficult to locate this transition accurately. Therefore we have chosen
a scheme that makes that the typical distance between pinned particles,
and therefore the size of the remaining liquid regions, is relatively
uniform. In practice we have generated an equilibrated liquid configuration
of $c\cdot N$ particles, at density $\rho=0.672$, using a molecular
dynamics simulation with periodic boundary conditions. Let us call a
typical configuration of this simulation $C_{\rm pin}$. Since this is
a liquid like configuration, all particles have a similar number of
nearest neighbors and also a well defined nearest neighbor distance
from each other. Subsequently we have scaled this configuration such
that the box size becomes the same as the one of configuration $C_{\rm
bulk}$. Then we have searched for each particles of the $C_{\rm pin}$
configuration the closest particle in the $C_{\rm bulk}$ configuration.
These closest particles have then been pinned permanently. In short this
procedure allows to select from the $C_{\rm bulk}$ configuration $c\cdot
N$ particles that can be pinned permanently and that have a relatively
unifom distance from each other.

Typically we have averaged our results over around 5 independent
realisations of the pinning field if $c$ was small and over up to 20
realisations when $c$ was large, i.e. around the Kauzmann point. The
total amount of CPU time used for obtaining the presented results is 
about 900,000 hours.

\section{Relaxation times}

\begin{figure}
\includegraphics[scale=0.35,angle=0]{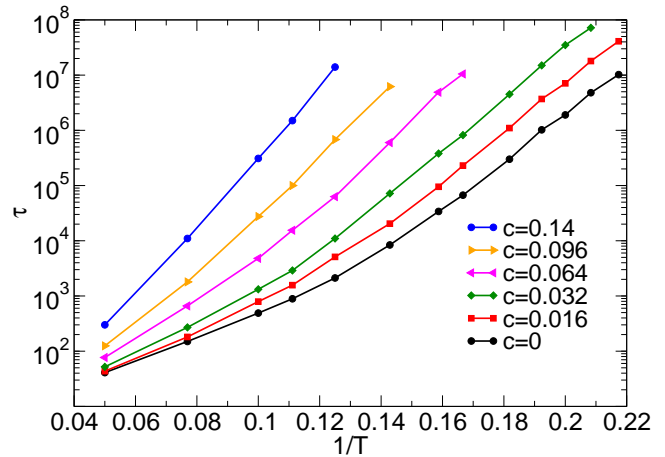}
\caption{Arrhenius plot of the relaxation time for different
value of the concentration $c$ of pinned particles.}
\label{fig1_sm}
\end{figure}

In order to determine the relaxation time of the pinned system as a
function of temperature and concentration $c$ we have used standard
molecular dynamics simulations with a system of $N=1000$ particles (step
size 0.05).  From these runs we have determined the self intermediate
scattering function $F_s(q,t)$, where $q$ is the wave-vector. Using for
$q$ the value of 5.5, which corresponds to the position of the maximum in
the static structure factor, we have defined the relaxation time $\tau$
via $F_s(q,\tau)=e^{-1}$. In Fig.~\ref{fig1_sm} we show the temperature
dependence of $\tau(T,c)$ in an Arrhenius plot. From this graph one
recognizes that for $c=0$, i.e. the bulk system, the data curves slightly
upwards, i.e. the system is a somewhat fragile glass-former,
as discussed in Ref.~\cite{berthier_09_S}. An increase
of $c$ has two effects: 1) The dynamics is slowing down very strongly
(by several orders of magnitude at intermediate and low $T$) and 
this slowing down becomes more pronounced if $T$ is decreased. 2) At the
highest $c$ considered the relaxation dynamics becomes Arrhenius-like,
i.e. the data falls on a straight line. This observation is in qualitative
agreement with the theoretical prediction by RFOT~\cite{cammarota_12a}
which shows that this aspect of the theory seems to be compatible with
the simulation data. (We mention that a similar decrease of fragility
with increasing $c$ has been observed in Ref.~\cite{kim_03_S}.)

Since Fig.~\ref{fig1_sm} demonstrates that the relaxation time
is a strong function of $c$, one can expect that the relevant
temperatures of the glass-former, such as the onset temperature
$T_0$, the mode-coupling temperature $T_c$, and the Kauzmann
temperature $T_K$ will also depend on $c$ and will increase if $c$
is increased. As a consequence it should become possible to cross the
line $T_K(c)$ by doing equilibrium simulations at low $T$ by increasing
$c$~\cite{berthier_12_S,cammarota_12a_S} and in the main text we
show that this crossing, and hence the investigation of the occurring
transition, can indeed be done in equilibrium.

\begin{figure}[ht!]
\includegraphics[scale=0.3,angle=0]{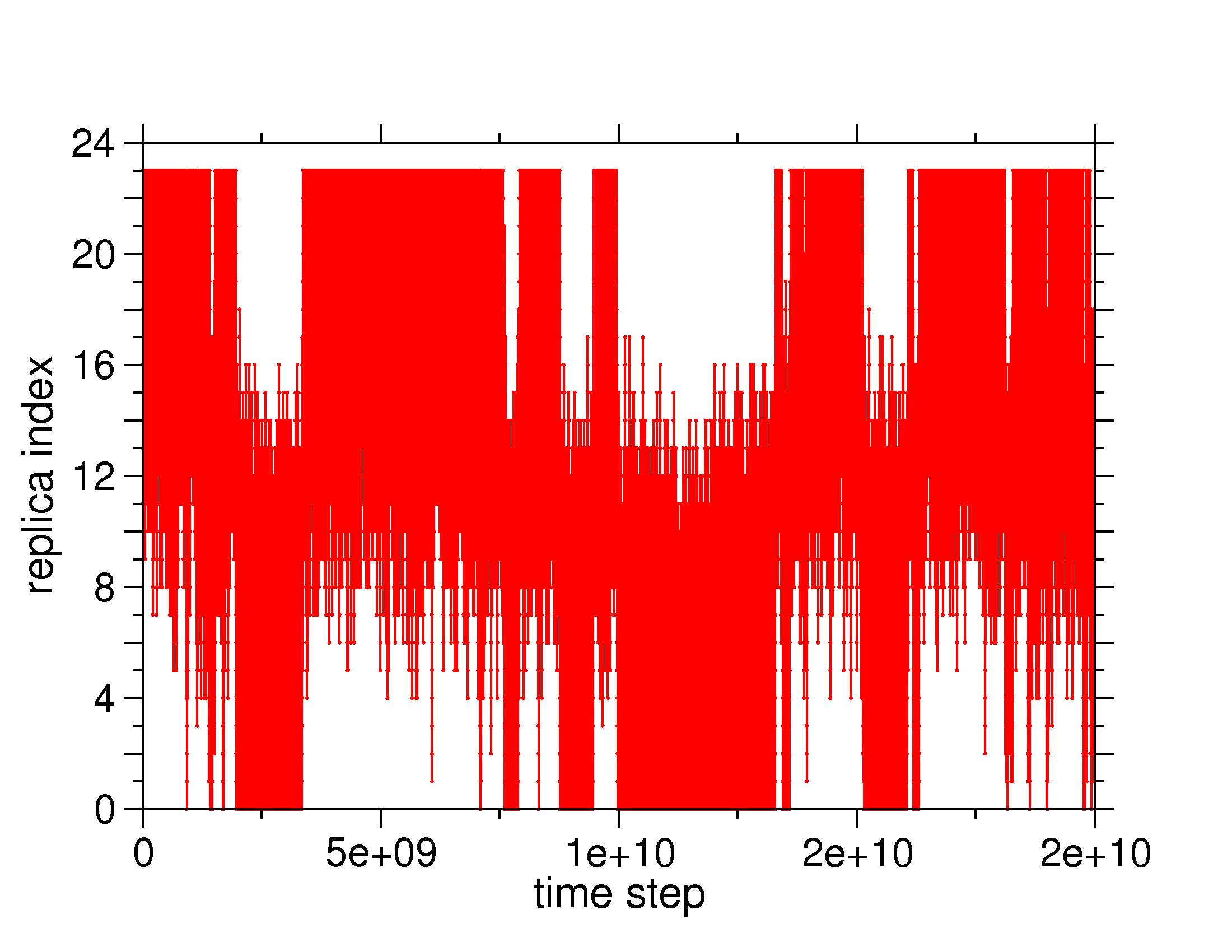}
\caption{Time dependence of a typical trajectory of a replica in
temperature space. The temperatures are labeled 0-23, with 0 being the
target temperature $T_{\rm liq}$. The example shown is for $T_{\rm
liq}=4.8$ and the concentration of pinned particles is $c=0.14$.}
\label{fig2_sm}
\end{figure}

Since the standard molecular dynamics slows down quickly upon
approaching the $T_K(c)$ line, we have used the parallel tempering
algorithm~\cite{hukushima_96_S,yamamoto_00_S}  using 24 replicas. The smallest
difference in temperature between two neighboring replicas was 0.3,
which has allowed for a good overlap of the two potential energy
distributions. Attempts to switch two neighboring replicas have been made
every 50,000 time steps.  We have checked that each parallel tempering run
has indeed reached equilibrium by following any given replica and making
sure that all temperatures have been sampled sufficiently.  A typical
path of a replica in temperature space is shown in Fig.~\ref{fig2_sm}
for our lowest studied temperature.

\begin{figure}[ht!]
\includegraphics[scale=0.3,angle=0]{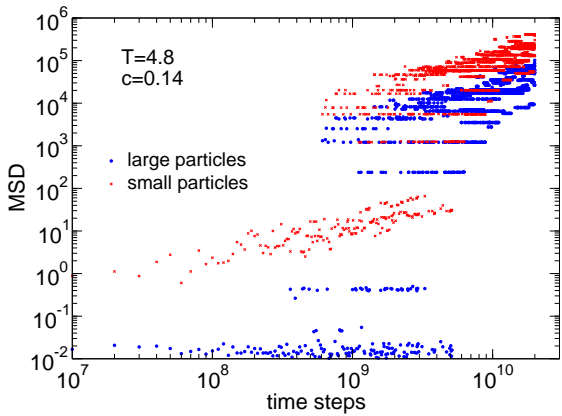}
\caption{Mean squared displacement of the particles as a function of time
using the parallel tempering dynamics.The circles and crosses correspond
the the large and small particles, respectively. $T=4.8$ and $c=0.14$.}
\label{fig3_sm}
\end{figure}

The parallel tempering algorithm does indeed allow to equilibrate the
system down to $T=4.8$ even if the concentration $c$ of pinned particles
is large. This can be recognized from Fig.~\ref{fig3_sm} where we show the
mean squared displacement (MSD) of the particles (distinguishing small
and large particles) as a function of time. To calculate the quantity
we have followed a given replica in temperature space and considered
only time intervals at which this replica was at the target temperature
$T_{\rm liq}$. The figure shows that at sufficiently long times the MSD
becomes very large, thus indicating that the particles do indeed move
through the box also at high values of $c$, even if their {\it relative}
arrangment does not change much, i.e. one is in the glass state. Or
stated otherwise: Each particle explores all the small patches shown in
Fig.~1a of the main text.

\end{document}